\documentclass{elsart}

\usepackage{amssymb}
\usepackage[dvips]{graphicx}

\journal{Theoretical Computer Science}

\begin{document}

\newcommand{\sdot}{{\scriptstyle \ldots}\;}

\begin{frontmatter}

  \title{Genetic Algorithms for the Imitation
  of Genomic Styles in Protein Backtranslation}

  \author{Andr\'es Moreira}

  \ead{anmoreir@dim.uchile.cl}

  \address{Center for Mathematical Modeling and Departamento 
           de Ingenier\'{\i}a Matem\'atica \\
           FCFM, U. de Chile, Casilla 170/3-Correo 3, Santiago, Chile}

  \begin{abstract}
    Several technological applications require the translation of a
	 protein into a nucleic acid that codes for it (``backtranslation'').
	 The degeneracy of
	 the genetic code makes this translation ambiguous; moreover, not
	 every translation is equally viable. The common answer to this problem
	 is the imitation of the codon usage of the target species. Here
	 we discuss several other features of coding sequences 
	 (``coding statistics'') that are relevant for the ``genomic style'' 
	 of different species.
	 A genetic algorithm is then used to obtain backtranslations that
	 mimic these styles, by minimizing the difference in the
	 coding statistics. Possible improvements and applications are discussed.
  \end{abstract}

\end{frontmatter}

\section{Introduction}

  The main components of the cell are nucleic acids (DNA and RNA) and proteins.
  Both are polymers, long words written in alphabets of 4 and 20 letters: 
  4 nucleotides for DNA and RNA, and 20 amino acids, for proteins. The 
  ``fundamental dogma'' of molecular biology describes the usual flow of information 
  in the cell, from DNA to mRNA to protein. The first step,
  {\em transcription}, preserves the sequence read from DNA, which is reversed 
  and complemented in the mRNA (in addition, the alphabet is slightly changed).
  It is straightforward to obtain the DNA from a given mRNA (it is called then
  {\em complementary} DNA, or cDNA); in fact, Nature does it: retrotranscription is
  performed by viruses and several small ``selfish'' units of information.

  The second step, {\em translation}, is more complicated: the mRNA is read,
  three nucleotides at a time, and an amino acid encoded by them is added to the
  forming protein, according to the well known genetic code (see Table 1). This
  nearly universal code associates to each triplet ({\em codon}) an amino acid,
  or the ``stop'' meaning.

\scriptsize
\begin{center}
{\small 
 Table 1: The (standard) Genetic Code} \nopagebreak \\
 \begin{tabular}{|cc|cc|cc|cc|cc|cc|cc|cc|}\hline
 aaa & K & aga & R & caa & Q & cga & R & gaa & E & gga & G & taa & stop & tga & stop \\
 aac & N & agc & S & cac & H & cgc & R & gac & D & ggc & G & tac &  Y   & tgc &  C   \\
 aag & K & agg & R & cag & Q & cgg & R & gag & E & ggg & G & tag & stop & tgg &  W   \\
 aat & N & agt & S & cat & H & cgt & R & gat & D & ggt & G & tat &  Y   & tgt &  C   \\
 aca & T & ata & I & cca & P & cta & L & gca & A & gta & V & tca &  S   & tta &  L   \\
 acc & T & atc & I & ccc & P & ctc & L & gcc & A & gtc & V & tcc &  S   & ttc &  F   \\
 acg & T & atg & M & ccg & P & ctg & L & gcg & A & gtg & V & tcg &  S   & ttg &  L   \\
 act & T & att & I & cct & P & ctt & L & gct & A & gtt & V & tct &  S   & ttt &  F   \\
 \hline
 \end{tabular}
 \end{center}
 \normalsize

  Unlike retrotranscription, the reversal of this second step (called 
  {\em backtranslation}) is ambiguous, due to the degeneracy of the genetic
  code: as can be seen in Table 1, amino acids are encoded by 1, 2, 3, 4 or 6
  different codons. 
  Backtranslation does not occur in natural systems\footnote{
  Though \cite{nashimoto01} suggests that it did occur at the origin of 
  life, and even proposes an {\em in vitro} device for backtranslation.},
  but is required for several purposes in genomics and biotechnology.
  The problem is not trivial, since different species have different
  ``genomic styles'' that determine which of the many preimages is used
  to code for a protein. Thus it may happen that we know the DNA for
  a given protein produced by, for instance, a plant, 
  but we want to synthesize the protein in a bacterium\cite{smith90}. We will need to 
  backtranslate the protein into the genomic style of this kind of 
  bacteria. In other cases, the protein is known but no DNA is known
  for it at all; this may happen with artificial proteins, or with
  proteins from unsequenced organisms.
  Other applications, like degenerate primers
  (for ``gene fishing'') and sequence analysis, will be discussed in
  the last section.
 
  The best known statistical feature of coding sequences is the
  presence of a periodicity of period 3, which is caused by the 
  structure of the genetic code and the asymmetry of the different
  codon positions\cite{herzel97,lee97}. This property
  is very important for distinguishing coding from non-coding sequences;
  however, it is not important for backtranslation, since it is
  shared by all organisms. On the other hand, we know that codon usage
  (the degree of preference for the different codons inside each synonymous class) 
  {\em does} distinguish one species from another; it is the best known feature of
  the different ``genomic styles''.
  
  The common approach to backtranslation relies on the imitation of the
  codon usage of the target species (the species whose style we want
  to imitate)\cite{pesole}. 
  This is the solution currently given by all commercial and non-commercial
  software, like GCG, EMBOSS, VectorNTI, EditSeq, AiO,
  and the online tools of Molecular Toolkit and Entelechon. 
  The only
  different approach we know is \cite{white98}, where a neural network
  was trained to perform backtranslation. However, it was done
  at the single amino acid level, and thus it cannot account for anything
  but codon usage. 

  This current solution can be improved; there are more features
  peculiar to the different coding styles\cite{guigo99,holste00}, which are
  in part or completely independent from codon usage\cite{guigo95}. 
  In the present article, we consider different possible statistics
  that may be associated to genomic styles, and then
  we apply a genetic algorithm to perform backtranslation, taking
  these features into account. Our approach considers DNA 
  {\em only as a symbolic sequence}, ignoring chemical properties or
  biological features. Furthermore, we will not use biological
  considerations to decide whether or not a statistical property
  needs to be imitated: we assume that any property distinguishing
  the style of a species must be considered in backtranslation (after
  all, in some cases the origin of known features remains obscure).
  All the statistics we consider were taken from
  the literature on sequence analysis, where their possible interpretations are
  discussed.

 \section{Notation, Materials}

   Let $A$=\{A, C, D, E, F, G, H, I, K, L, M, N, P, Q, R, S, T, V, W, Y\}
	and $B$=\{$a$,$c$,$g$,$t$\} be the alphabets for amino acids and nucleotides,
	respectively, and denote $B^{3*}=(B^3)^*$.
   Let $\tau:B^{3*}\rightarrow (A\cup \{stop\})^*$ be the translation
	of a sequence according to the genetic code. In fact, $\tau$ may
	depend on small variations to the code which do occur in some species
	and organelles; however, here we will assume the code to be universal.
   Furthermore, we will consider the sequences
	{\em without the start and stop} signals, i.e., cutting the $atg$
	codon that initiates a protein and the stop codon that marks its
	end.
	
	We will say that
	a function (or stochastic procedure) $\beta:A^*\rightarrow B^{3*}$
	is a {\em guess} iff $\tau\circ \beta = id_{A^*}$. If 
	$C \subset A^*$, we will denote $\beta(C)=\{\beta(u):u\in C\}$.	
	A particular guess that will be used for comparison purposes is 
	the canonical backtranslation procedure, which backtranslates each
	amino acid using the empirical frequencies of its codons as
	probabilities; we will denote it as $\beta^{cu}_{sp}$, with the 
	subindex indicating the species whose codon usage table was used.

   Given a sequence $w\in B^{3*}$, $w=w_0,w_1,\ldots$ and $i=1,2,3$, we will talk about 
	the letters in {\em codon position i} to refer to $w_{i-1}$, $w_{i+2}$, $w_{i+5}$,
	\ldots. We will denote with $\pi_{ry}$, $\pi_{ws}$ and $\pi_{mk}$ the three
	most usual projections of $B$ into $\{0,1\}$, as follows.
	We will use the same symbols to refer to the extensions of these
	functions to $B^{3*}$ (projecting each letter).

\begin{center}
 \begin{tabular}{|c|cccc|c|}\hline
             & $a$ & $c$ & $g$ & $t$ & refers to: \\ \hline
  $\pi_{ry}$ &  0  &  1  &  0  &  1  & purine/pyrimidine \\
  $\pi_{ws}$ &  0  &  1  &  1  &  0  & weak/strong \\
  $\pi_{mk}$ &  0  &  0  &  1  &  1  & amino/keto \\
    \hline
 \end{tabular}
 \end{center}

	It is important to notice that many characters in $\beta(u)$ are almost
	or completely determined by $u$. Amino acid
	K, for instance, is coded by $aaa$ and $aag$; the first and the second
	position will be $a$ in any backtranslation, and the third one
	will be either $a$ or $g$ (and will have $\pi_{ry}=0$, so that for any $\beta$,
	$\pi_{ry}(\beta(K))=001$).
	The next table shows the number of amino acids for which characters are fixed 
	in the different codon positions for the different binary alphabets. Most of
	the ambiguity of backtranslation is in the third position.

\small
\begin{center}
 \begin{tabular}{|c|ccc|}\hline

              & $\pi_{ry}$ & $\pi_{ws}$ & $\pi_{mk}$ \\ \hline
  Cod. Pos. 1 &      18    &      18    &     18     \\ 
  Cod. Pos. 2 &      19    &      20    &     19     \\ 
  Cod. Pos. 3 &      11    &       2    &      2     \\ 
  \hline
 \end{tabular}
 \end{center}
 \normalsize

  {\bf Materials } 

  We extracted coding sequences from Genbank\cite{genbank} release 131 (August
  2002), belonging to the following species: Methanosarcina acetivorans C2A ($A1$),
  Sulfolobus solfataricus ($A2$), Escherichia coli ($B1$), Bacillus subtilis ($B2$),
  Streptomyces coelicolor A3(2) ($B3$), Mesorhizobium loti ($B4$), Nostoc sp. PCC 7120
  ($B5$), Saccharomyces cerevisiae ($E1$), Arabidopsis thaliana ($E2$), Drosophila melanogaster
  ($E3$), Caenorhabditis elegans ($E4$) and Homo sapiens ($E5$). The selection of
  species was done trying to have abundant sequences and a rather good representation of the
  tree of life. All coding sequences (``CDS'' features in Genbank) were extracted,
  provided that they were complete, univoque, and longer than 1029 nucleotides. The average
  length of the sequences varies between 1500 for $A1$ and 2456 for $E3$.
  Please notice that introns -intervening sequences- were removed from the sequences; 
  this may affect the coding statistics that depend on relations between
  distant nucleotides. 
  We will use the abbreviation of a species to
  refer to the set of its coding sequences, or to the set of the corresponding
  proteins, depending on the context. Thus, an expression
  like $\beta^{cu}_{B1}(E5)$ denotes a set of backtranslations obtained
  for all proteins encoded by the coding sequences of $E5$, 
  obtained by the standard backtranslation method, considering the codon usage 
  of $B1$.

 \section{Coding Statistics}

  Here we discuss the results of computations performed
  on our set of species for several features that have been studied
  in coding sequences, ``generally known as coding statistics,
  since their behavior is statistically distinct on coding and
  non-coding regions''\cite{guigo95}. Discussions about the most common
  coding statistics, their relations, and their use for gene finding,
  can be found in \cite{guigo99} and \cite{holste00}.  However, we
  are not interested in the difference between coding and 
  non-coding regions; rather, we want those statistics 
  that contribute to the ``genomic style'' of a species.
  
  The notion of genomic style has been around since the
  ``genome hypothesis'' of Grantham \cite{grantham80,grantham86}, who first
  recognized the idiosyncratic nature of codon usage. Later,
  Karlin used the bias in dinucleotide usage
  as the ``genomic signature'' of a species \cite{karlin97}. 
  Forsdyke suggests that the species ``broadcast'' their genes
  in different $g+c$ frequencies \cite{forsdyke96}, and that
  this could be crucial for speciation; in this way,
  genomic styles could be the first line of an immune system\footnote{Indeed,
  \cite{cristillo01} shows that some viruses
  may mimic the genomic style of their host, in order
  to be expressed.}.
  There
  have been other proposals, usually for phylogenetic purposes.
  The reasons for the existence of different styles are debatable:
  for instance,
  changes in the molecular machinery, tRNA abundance, environmental 
  temperature, different biases in the mutation rates, 
  the requirements of messages other than the protein
  sequences\cite{trifo98}, etc. The exact 
  causal relations are subject to discussion.

  In order to improve the profile of genomic styles, we want to 
  choose those statistics which: 
  (1) have typical and statistically sound values for each species, 
      with small variability, 
  (2) have different values in different species, and 
  (3) do not depend (exclusively) on the amino acids encoded 
      by a sequence (i.e., they {\em do} depend on backtranslation).
  Because of space limitations, we will not give the values
  of all computations; in the graphics, not all the species will be
  displayed, if it is not required. 
  Moreover, we will dispense from data in the case of well
  known facts. All computations and data sets can be found at \cite{web}.

  \subsection{Nucleotide frequencies}

	The most natural computation is the frequency of the 
	four nucleotides in the sequences, as well as their frequencies
	in the different codon positions. For each sequence $w\in B^{3*}$,
	$w=w_0,\ldots,w_{3N-1}$, and each nucleotide $\alpha$, we	compute
	\[
	 \rho_\alpha(w) \;=\; \frac{1}{3N} \sum_{i=0}^{3N-1} \delta_\alpha(w_i)
	 \quad , \quad
	 \rho_\alpha^j(w) \;=\; \frac{1}{N} \sum_{i=0}^{N-1} \delta_\alpha(w_{3i+j-1}),
	 j=1,2,3
	\]
	where $\delta_\alpha(x)$ is $1$ if $x=\alpha$ and 0 otherwise.
	Our computations confirm a number of facts already known in the
	literature, like ``Chargaff's second law'', which states
	that $\rho_a \approx \rho_t$ and $\rho_c \approx \rho_g$ as can
	be observed in Graphic 1a. Since, in addition, $\sum_{\alpha\in B}\rho_\alpha=1$,
	Chargaff's law implies positive correlation between complementary nucleotides
	($a$ with $t$, and $c$ with $g$) and negative correlation between 
	non-complementary ones. 
	Thus we can reduce the study to a single value; the usual choice is
	$\rho_{g+c}=\rho_c+\rho_g$.
	It is well known that $\rho_{g+c}$ has different values in different species,
	and that all the genes in a species have similar values; this can be
	seen in Graphic 1b, with histograms showing the number of sequences of each
	species in different $\rho_{g+c}$ ranges. 
	Some qualifications are due: First, it is also known 
	that eukaryotic genomes are
	organized in large ``islands'' called {\em isochores} \cite{macaya76}, with 
	different $\rho_{g+c}$ values but each of them relatively homogeneous.
	Moreover, in a set of closely related species $\rho_{g+c}$ may depend more
	on the genes than on the species\cite{ma02}. However, the general pattern holds,
	and it is used both for the detection of genes (since genes tend to be
	$\rho_{g+c}$-richer than non-coding regions) and in the detection of horizontally
	transferred genes (see section \ref{sec:lateral}).

 \begin{figure}[hbt!]
  \begin{center}
   \includegraphics[width=10.5pc]{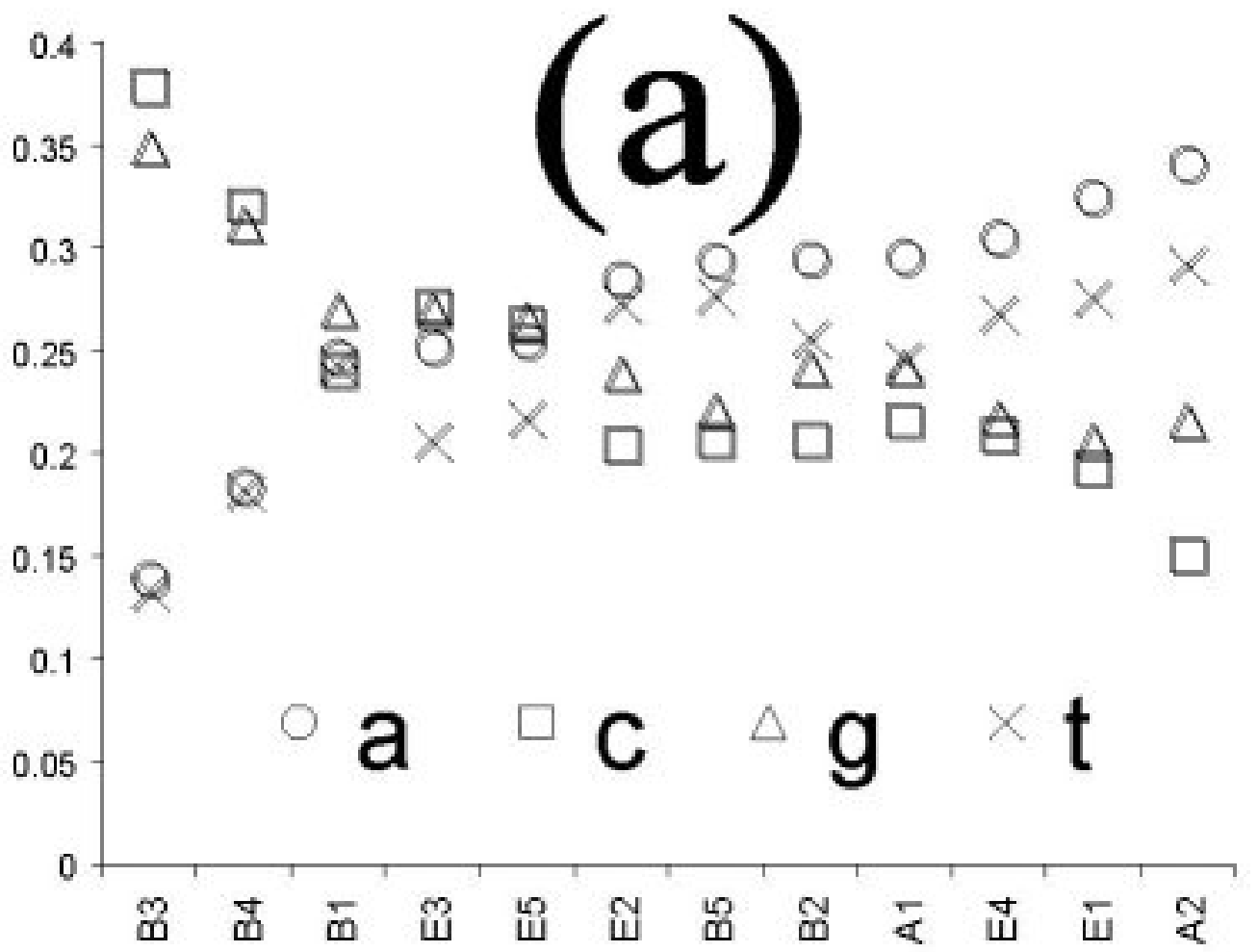}
   \includegraphics[width=10.5pc]{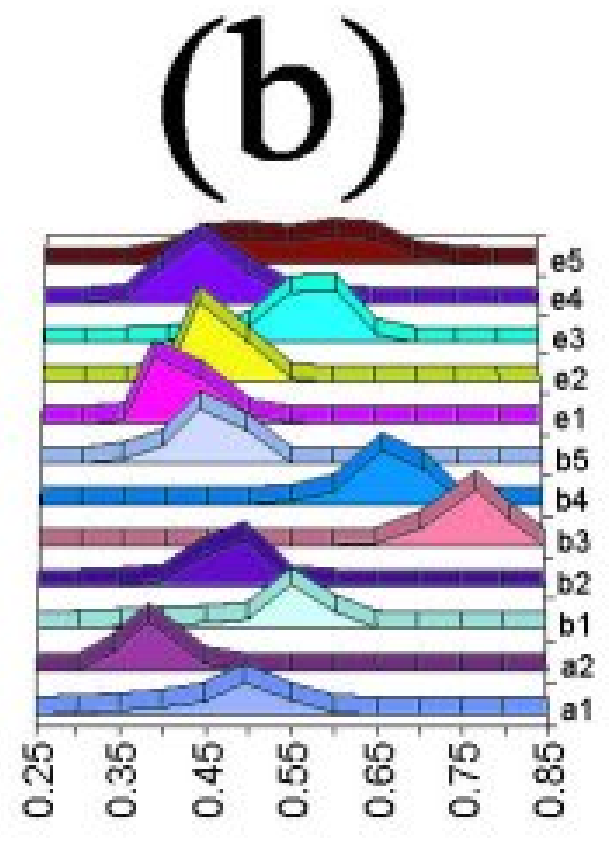}
   \includegraphics[width=10.5pc]{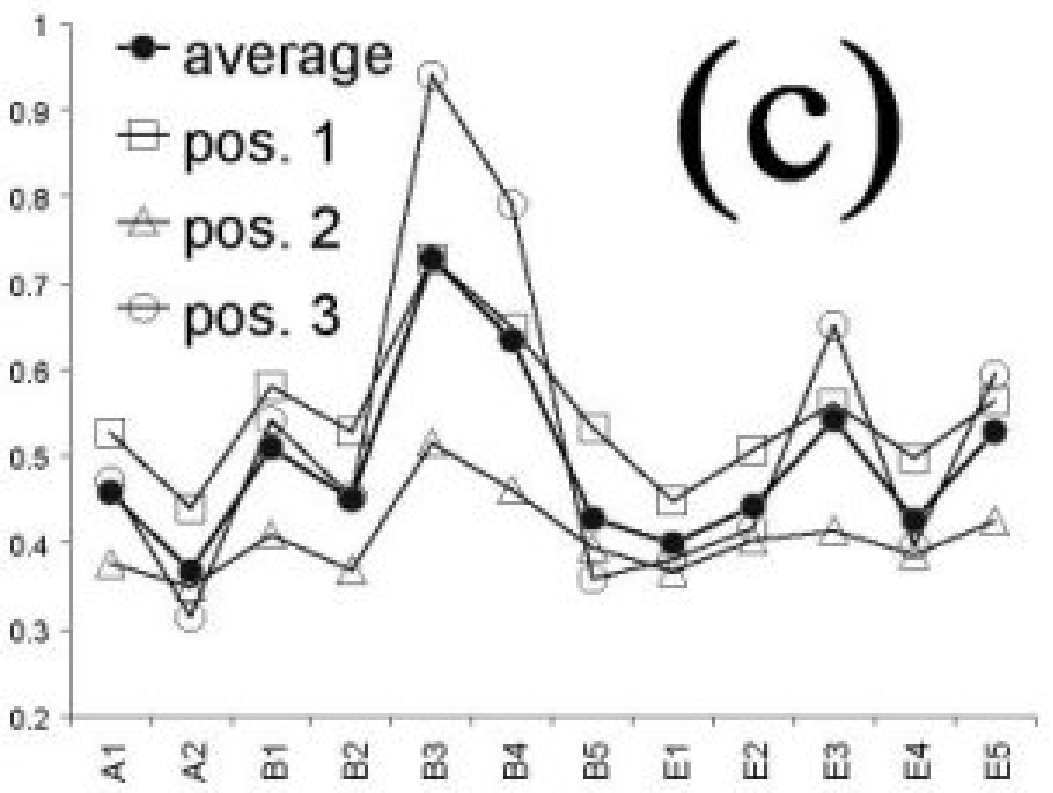}
  \end{center}
 \caption{(a) Nucleotide frequencies. (b) Histograms for $\rho_{g+c}$. (c) $\rho_{g+c}$ in different
 codon positions.}
 \label{fig:gr1}
 \end{figure}

	Graphic 1c shows the values of $\rho_{g+c}^j=\rho_c^j+\rho_g^j$ for the different
	species, together with $\rho_{g+c}$. We notice the existence of wide variations 
	in the $\rho_{g+c}$ composition
	depending on the codon position. In addition, 
	extreme values of $\rho_{g+c}$ are usually supported by extreme values
	of $\rho_{g+c}^3$; this shows that the sequences were adapted to get a certain
	$\rho_{g+c}$ level, and that the third -usually synonymous- codon position
	was used for this purpose. As can be seen in Table 2, $\rho_{g+c}^1$ and 
	$\rho_{g+c}^2$ are almost entirely determined by the encoded amino acids.

  \subsection{Codon usage}

   The frequency of a given codon $C=c_0,c_1,c_2\in B^3$ in a sequence
	$w=w_0,\ldots,w_{3N-1} \in B^{3*}$ is defined as
	$
	 \frac{1}{N} \sum_{i=0}^{N-1} 
	 \delta_{c_0}(w_{3i}) \delta_{c_1}(w_{3i+1}) \delta_{c_2}(w_{3i+2})  
	$.
	For each codon $C\in B^3$, we define its synonymous class 
	$\theta(C)=\{C'\in B^3\; : \; \tau(C)=\tau(C')\}$. Then the {\em synonymous
	codon usage} and the {\em relative synonymous codon usage} \cite{sharp87} of $C$ 
	are defined as
	\[
	 SCU_C \;=\; \frac{\rho_C}{\displaystyle \sum_{C'\in \theta(C)} \rho_C'}
	 \quad , \quad
	 RSCU_C \;=\; \frac{|\theta(C)| \, \rho_C}{\displaystyle \sum_{C'\in \theta(C)} \rho_C'}
	        \;=\; |\theta(C)| \, SCU_C 
	\]

   As we mentioned above, the codon choice pattern was noted very early
	to be a signature of the species, and our data confirm 
	this. We will dispense with extensive SCU tables, since they are well known
	in the literature, and available in public databases\cite{nakamura99}.
	As we said before, the common approach to backtranslation 
	uses SCU as the probability of choosing a certain codon, given the amino acid. RSCU
	is used for comparisons between codons from different synonymous classes.

  \subsection{Dinucleotides}

	Most pu\-bli\-shed re\-sults on di\-nucleo\-ti\-de fre\-quen\-cies 
	con\-si\-der long DNA sequences,
	including both coding and non-coding regions \cite{burge92,haring99,shioiri01}. Our
	own computations, in spite of being limited to coding sequences, confirm most 
	of the facts already noted	by the different authors. This
	accounts for the fact that dinucleotide frequencies are not considered
	as ``coding statistics'': their behavior is similar in coding and in non-coding
	sequences. However, they do exhibit characteristic patterns according to the
	different species and groups. Karlin \cite{karlin97} even used them to
	define the {\em genome signature} of a species as the collection 
	$\{\varrho^*_{\alpha\beta}\}$, with $\alpha$ and $\beta$ ranging over $B$. 
	Here $\varrho_{\alpha\beta}=\rho_{\alpha\beta}/\rho_\alpha \rho_\beta$
	(with $\rho_{\alpha\beta}$ being the frequency of the dinucleotide $\alpha\beta$)
	and $\varrho^*$ is the computation of $\varrho$ over the sequence concatenated
	to its inverse complement (in order to get the information about both
	DNA strands).

   {\bf IDH. }
	There is an	interesting set of indices which can be computed from
	dinucleotide frequencies. The so called {\em index of DNA homogeneity} 
	(IDH) was proposed by Miramontes et al \cite{mira95}
	and is defined for a binary sequence as
	$d = \frac{ \rho_{00} \rho_{11} - \rho_{01} \rho_{10} }{ \rho_0 \rho_1 }$.
	We define $d_{ry}(w)=d(\pi_{ry}(w))$, $d_{ws}(w)=d(\pi_{ws}(w))$, 
	and $d_{mk}(w)=d(\pi_{mk}(w))$. This index expresses the degree of local
	homogeneity of the sequence: long stretches of 0 or 1 will cause
	$d$ to be near 1, while strong alternation will push it toward -1. The
	three indices $d_{ry}$, $d_{ws}$ and $d_{mk}$ are not independent,
	and since $\pi_{mk}$ is the least meaningful of the binary
	projections, the choice in \cite{mira95} was to plot the species
	in the $(d_{ry},d_{ws})$ plane. The corresponding map with our own
	data is in Graphic 2a. Graphic 2b displays the distribution
	of the values in the sequences of some species.
	Both the specificity and the classificatory power of IDH
	can be clearly noted.

 \begin{figure}[hbt!]
  \begin{center}
   \includegraphics[width=15pc]{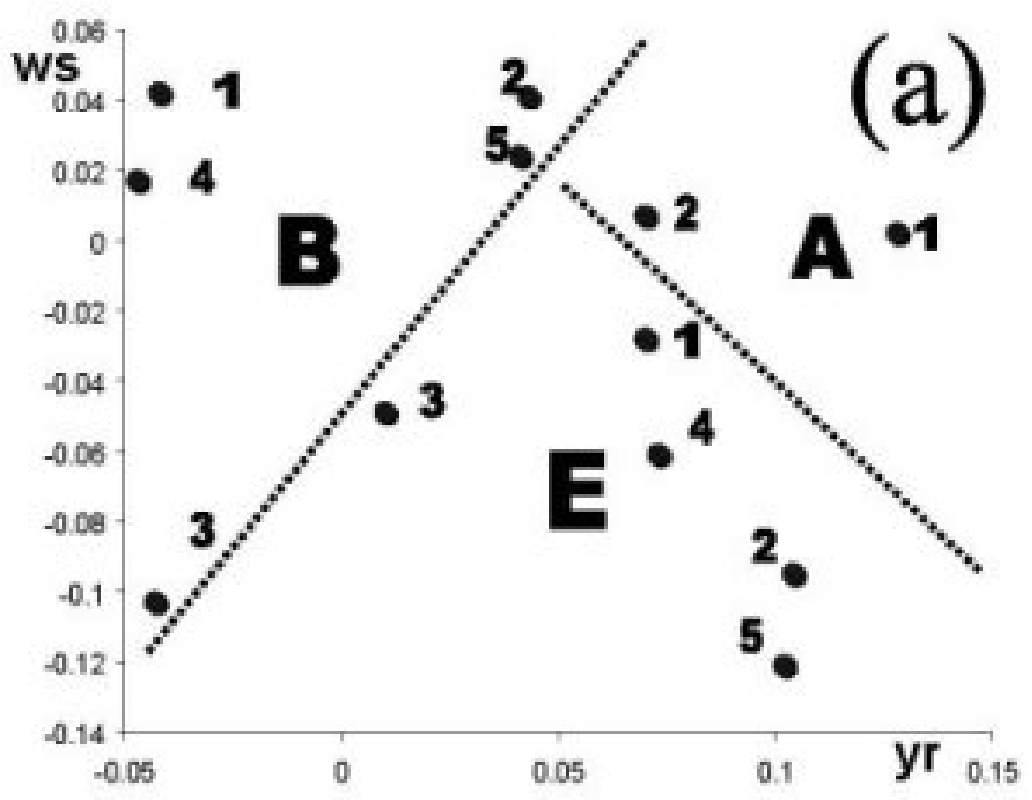}
	$\quad$
   \includegraphics[width=15pc]{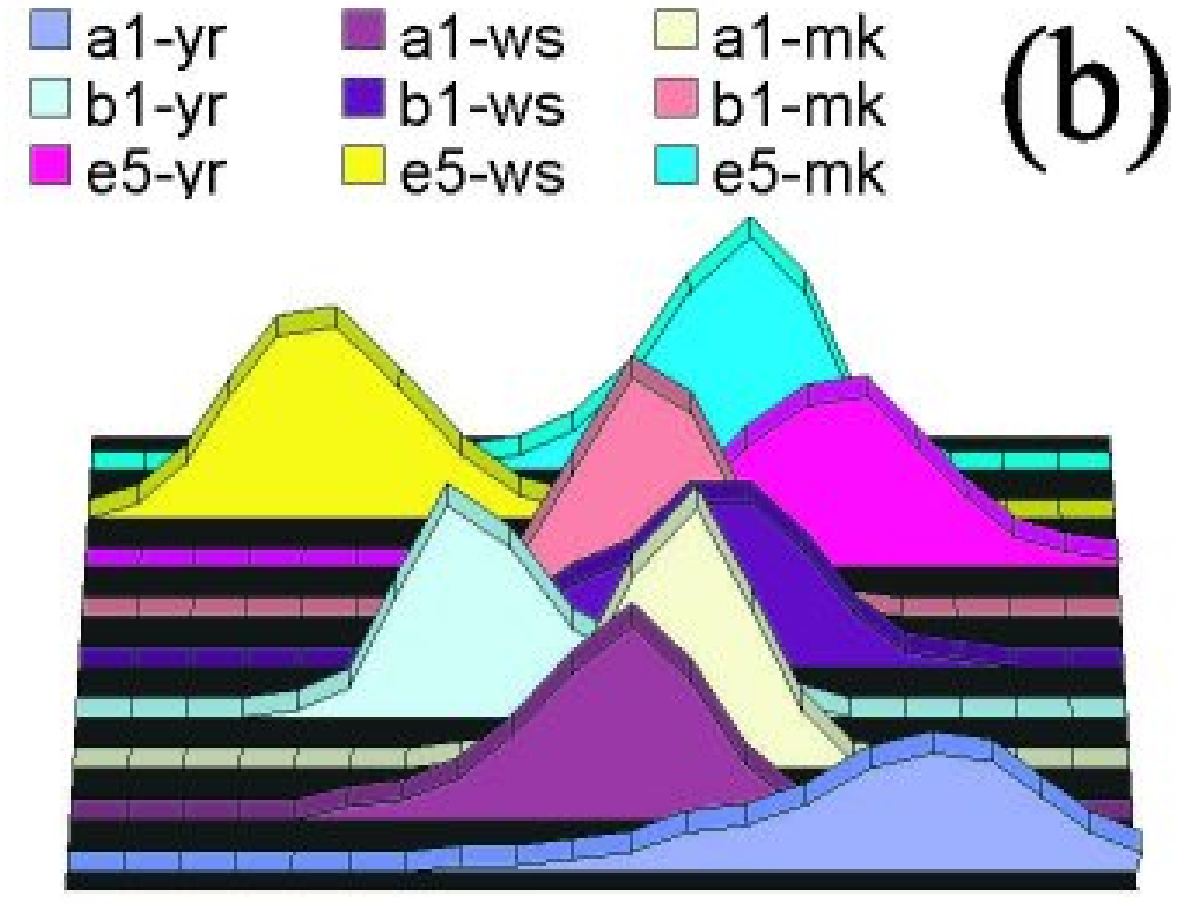}
  \end{center}
 \caption{(a) Position of species in the $(d_{ry},d_{ws})$ plane.
          (b) Histograms for IDH in some species.}
 \label{fig:gr2}
 \end{figure}

  \subsection{Fourier harmonics and Periodicities}

	Another common tool for DNA analysis is the discrete Fourier transform\cite{lobzin00}.
	For a binary sequence $w=w_0,\ldots,w_{N-1}$, we define
	the spectrum and its $m$-smoothed version:
	\[
	 S_n(w) \;=\; \frac{1}{N^2} \left| \sum_{k=0}^{N-1} w_k \, e^\frac{2\,\pi\, i\, n\, k}{N} \right|^2
	\quad \textrm{ , } \quad
	\tilde{S}^m_n(w) \;=\;  \sum_{k=n-m}^{n+m} \frac{S_k(w)}{2m+1}
	\]
	$S_n(w)$ measures the frequency content of `frequency' $n$, which corresponds
	to a period $\frac{N}{n}$; the smoothed value helps to remove the dispersion
	that appears for small data sets.

	The main and better known periodicity in DNA sequences is of period 3; it
	can be explained by the asymmetry in the codon positions \cite{herzel97,lee97},
	though its presence in tRNA genes suggests some other origin. Another well 
	documented periodicity is of period 10.5 $\pm$ 0.5; it has been
	attributed to requirements from the structure of both DNA and proteins, and
	the exact contribution of each is unclear. Some periodicities of higher
	periods have been shown, but they
	are not statistically significant for the typical lengths of genes.
	
 \begin{figure}[hbt!]
  \begin{center}
   \includegraphics[width=10.5pc]{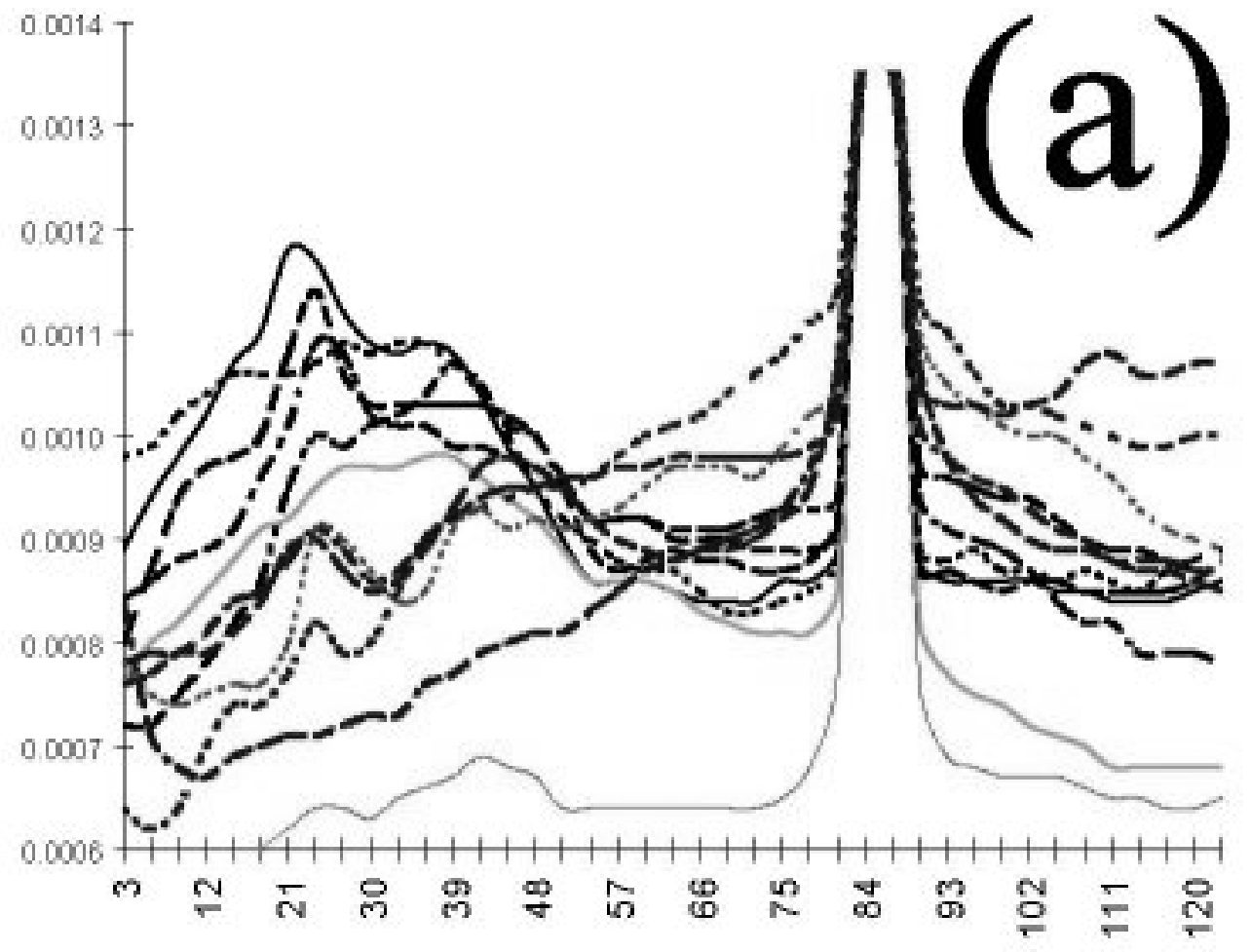}
   \includegraphics[width=10.5pc]{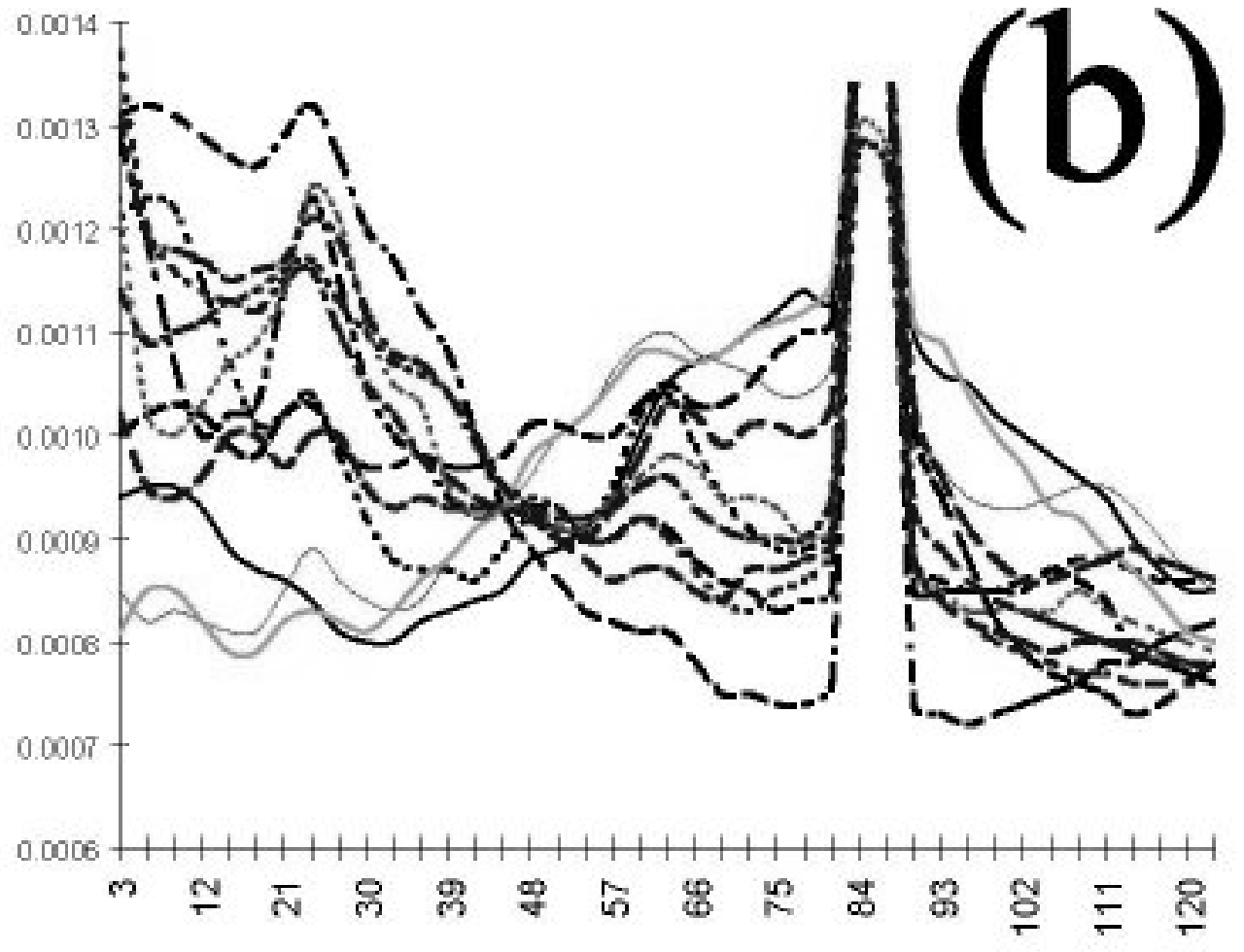}
   \includegraphics[width=10.5pc]{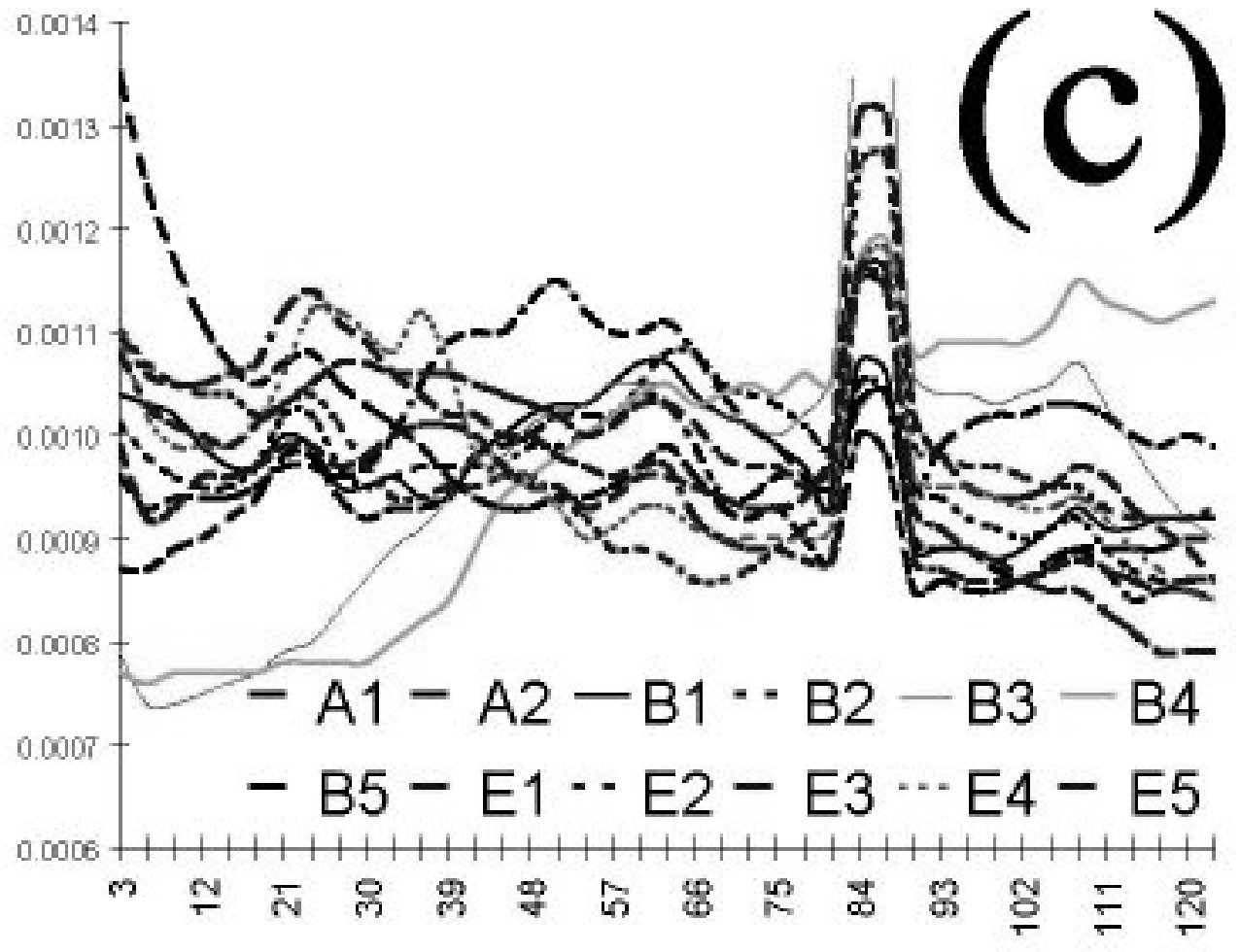}
  \end{center}
 \caption{$\tilde{S}^5$ for (a) $\pi_{ws}$, (b) $\pi_{ry}$ and (c) $\pi_{mk}$.}
 \label{fig:gr3}
 \end{figure}

	We divided each sequence in non-overlapping windows of length 256,
	and used the fast Fourier transform (FFT) algorithm to compute
	$\tilde{S}^5 \circ \pi_{ry}$, $\tilde{S}^5 \circ \pi_{ws}$
	and $\tilde{S}^5 \circ \pi_{mk}$ for all the species. The
	results were averaged and are shown in Graphics 3a, 3b and 3c for some
	of the species; only part of the ordinate axis is used, in order to highlight
	their differences.
	The two periodicities mentioned before are present: there is a big peak
	at $n=85$ for the three projections in almost all the species (the top of the
	peaks is outside the graphics); this corresponds to
	a period of $\frac{256}{85}\approx 3$. There is also a minor peak
	around $n=24$, present for most species and for most projections, corresponding
	to the period $\frac{256}{24}\approx 10.5$; there are some differences between
	species, a fact that has been observed before and is related to the
	various origins of this periodicity.
	
	To show the specificity of the spectrum, we chose a set of 20 collections
	of sequences, each set selected at random to be 1\% of $E5$. We computed
	the average of spectra for each set; the results for $\pi_{ws}$
	are shown in Graphic 4a.

  {\bf Position dependent spectra.} To take into account the asymmetry of
  the different codon positions, we computed the spectra for the three 
  subsequences $w^{(i)}_n=w_{3n+i}$, $i=0,1,2$, using windows of length
  64 (data not shown). In absence of period 3, the most notorious feature
  is a peak at $n=18$, corresponding to a period $\frac{64}{18}\approx 3.5$
  in the subsequence, and hence 10.5 in the sequences; it is by far stronger
  for the middle codon position, 
  a fact that hints for dependence on the amino acid sequence.

 \begin{figure}[hbt!]
  \begin{center}
   \includegraphics[width=15pc]{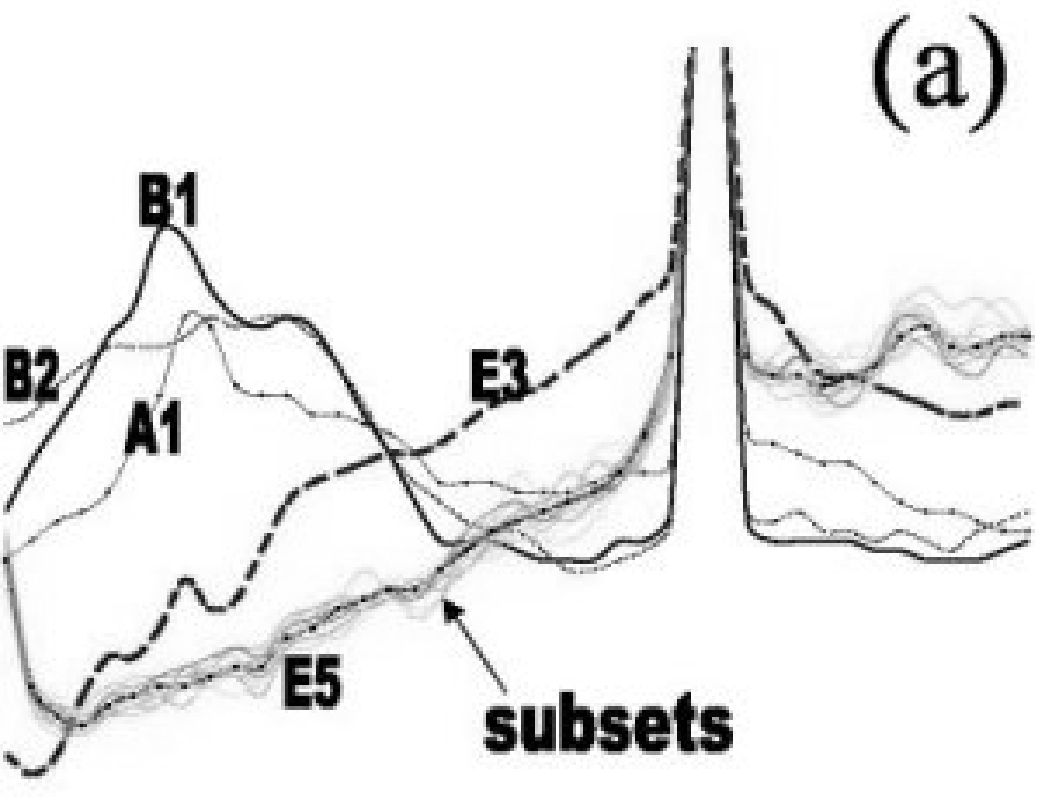}
   \includegraphics[width=15pc]{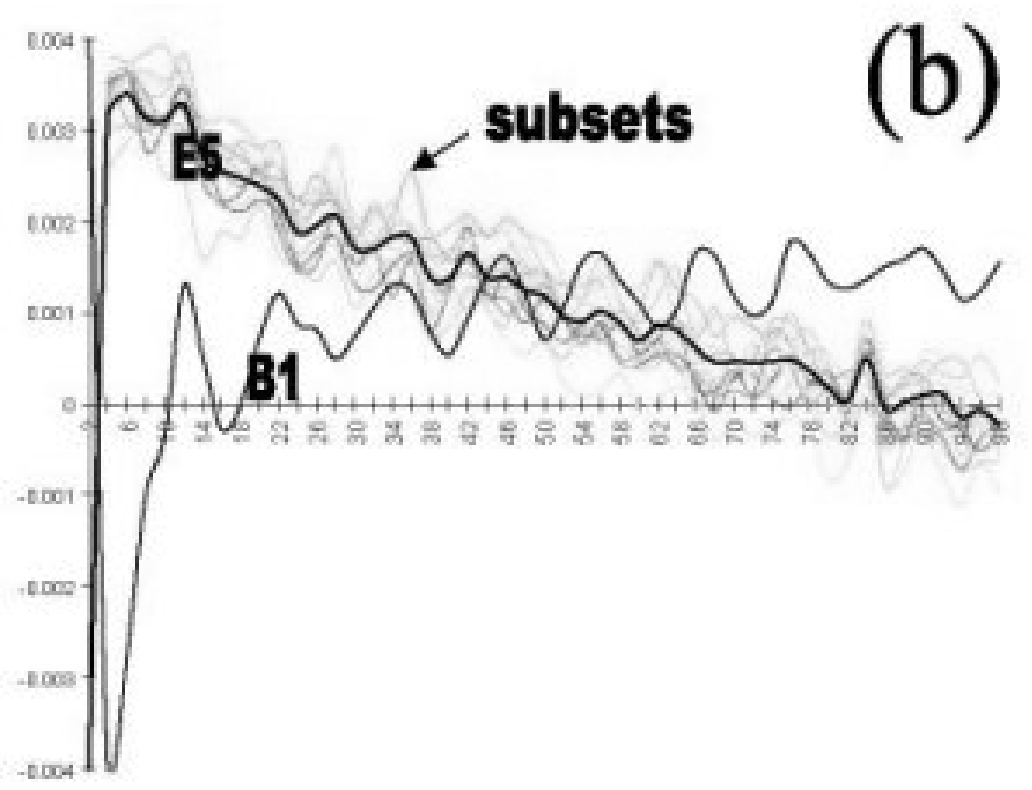}
  \end{center}
 \caption{Dispersion of (a) $\tilde{S}^5$ and (b) $\tilde{\Gamma}$ for $\pi_{ws}$ in $E5$.}
 \label{fig:gr4}
 \end{figure}

  \subsection{Autocorrelation functions}

  Correlation functions \cite{herzel95,herzel98} measure the excess or defect of nucleotides at different
  distances; if $\rho_{\alpha,\beta}(d)$ is the frequency with which we find a `$\beta$'
  $d$ positions after a `$\alpha$', then what we compute is $\rho_{\alpha,\beta}-\rho_\alpha \rho_\beta$.
  More precisely, what we compute for a sequence $w=w_0,\ldots,w_{N-1}$ is
  \[
   \Gamma_{\alpha,\beta}(d)[w] \;=\; \frac{1}{N-d} \sum_{i=0}^{N-d-1} \delta_\alpha(w_i) \delta_\beta(w_{i+d})
	\; - \; \rho_\alpha(w) \rho_\beta(w)
  \]
  We computed $\Gamma_{0,0}$ for $\pi_{ry}$, $\pi_{ws}$, $\pi_{mk}$.
  The most notorious result of this computation is the strong oscillation due to period 3;
  this can be removed by considering the smoothed version, $\tilde{\Gamma}_{\alpha\beta}(d)=\frac{1}{3}
  \sum_{i=d-1}^{d+1} \Gamma_{\alpha\beta}(i)$; when this was done, the periodicity of
  period 10.5 could also be seen. 
  To give an idea of the shape of the curves, and to
  show their specificity, Graphic 4b shows the results for $\pi_{ws}$, for $B1$, $E5$, and the same
  subsets of $E$ used in Graphic 4a. In general, $\Gamma$ behaves very similar to the Fourier 
  transform, in specificity and in the dependencies on alphabet and/or codon position. This is
  no surprising, since both express the same information (if $\Gamma$ is computed for a circular
  sequence, then it can be recovered form the spectra, and vice versa, by the Wiener-Khinchin
  theorem). Position dependent autocorrelation functions were also computed, with no
  unexpected results.

 \section{Backtranslation strategy}

 \subsection{Genomic style beyond codon usage}
 \label{sec:relaciones}

	We will consider all of the coding statistics reviewed in the previous section
	as features defining the genomic style of a species. It is important to notice
	that they are not (or not directly) dependent on the codon usage; if this
	were the case, then genomic style would reduce to RSCU, and the current approach
	to backtranslation would be already optimal. 

	It is clear that $\rho_\alpha$ and $\rho^j_\alpha$ are
	recovered by RSCU, if the amino acid composition is kept constant (this is the case
	in $\beta^{cu}_{B1}(B1)$ and $\beta^{cu}_{E5}(E5)$); in general, since amino acid composition
	is rather similar in all the different species (data not shown), we can expect
	nucleotide frequencies to be conserved.

	For dinucleotides, this is not so clear, even if the amino acid frequencies
	are kept: in spite of recovering the number of dinucleotides starting at the
	first and second codon positions, RSCU will not recover those starting at the
	third. This is important, since most of the degeneracy is in this position,
	and ``genomic style'' depends strongly on it; moreover, mutation rates tend
	to be affected by the neighboring nucleotides \cite{arndt02,hess94}, 
	in ways that are species-dependent.
	In particular, Miramontes et al \cite{mira95} show that their
	indices (IDH) are {\em not} determined by codon usage, even when the amino acid
	frequency was conserved. Our data (not shown) confirm it.

	As for the Fourier spectra, Guig\'o \cite{guigo95,guigo99} shows that it is
	rather independent from $\rho_{g+c}$. 
	To discard dependence on RSCU, we computed the spectra on $\beta^{cu}_{B1}(B1)$, 
	$\beta^{cu}_{B1}(E5)$, $\beta^{cu}_{E5}(E5)$ and $\beta^{cu}_{E5}(B1)$; results
	for $\tilde{S}^5_n\circ \pi_{ws}$ are displayed in Graphic 5a.
	We can see that all the sets of guesses lie between the real spectra, with
	codon usage being a bit more relevant than the amino acid sequences (the species);
	this was also the case for $\pi_{ry}$ and $\pi_{mk}$ (data not shown).
	Although the autocorrelation function contains the same information as the spectrum, 
	the details of each one are the main lines of the other, and thus, each may be
	considered apart. Graphic 5b displays computations of 
	$\tilde{\Gamma}_{0,0}\circ \pi_{ws}$
	over the same sets; it can be noticed that in this case the species
	(amino acid sequences) are the major contribution, with only a small effect of RSCU.
 \begin{figure}[hbt!]
  \begin{center}
   \includegraphics[width=15pc]{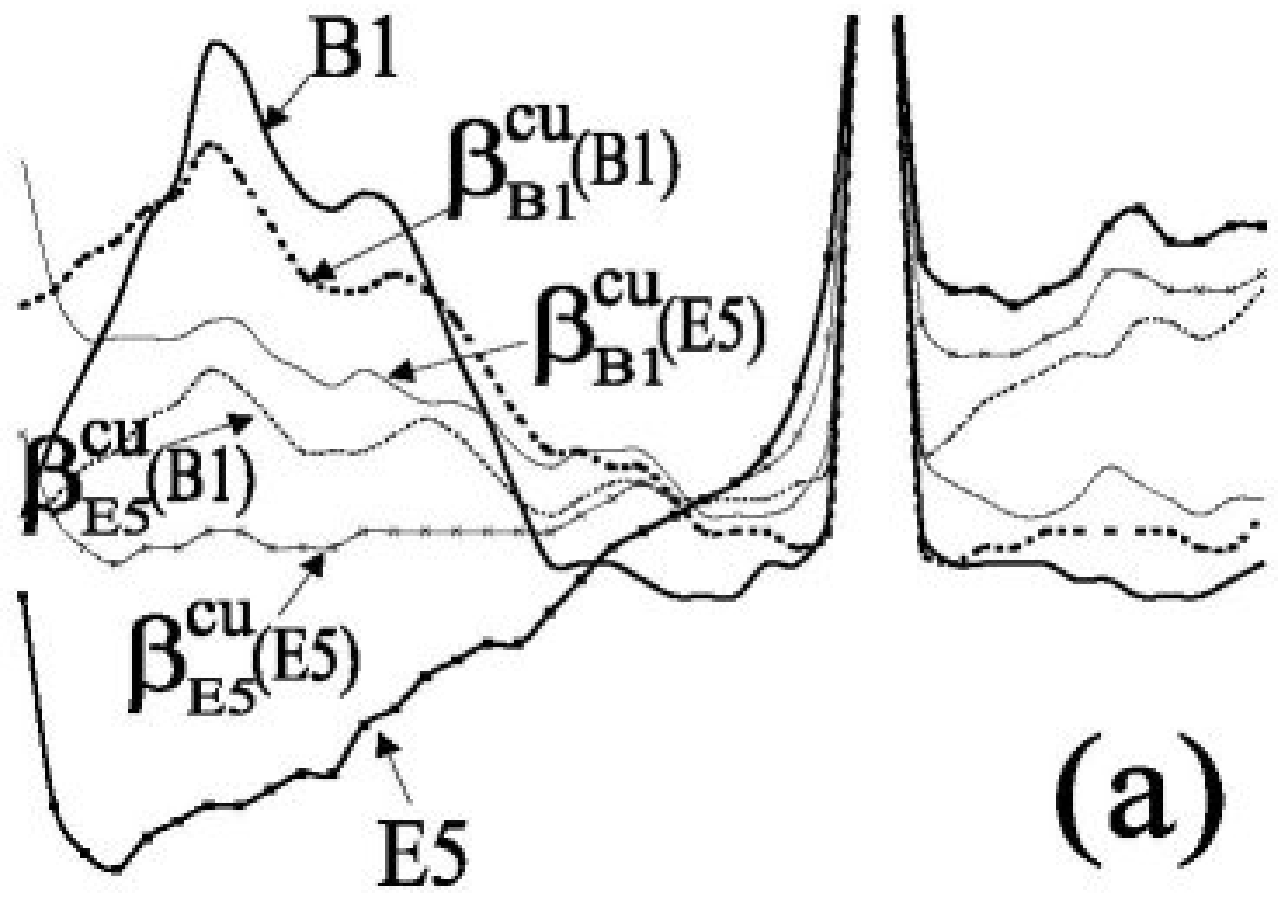}
   \includegraphics[width=15pc]{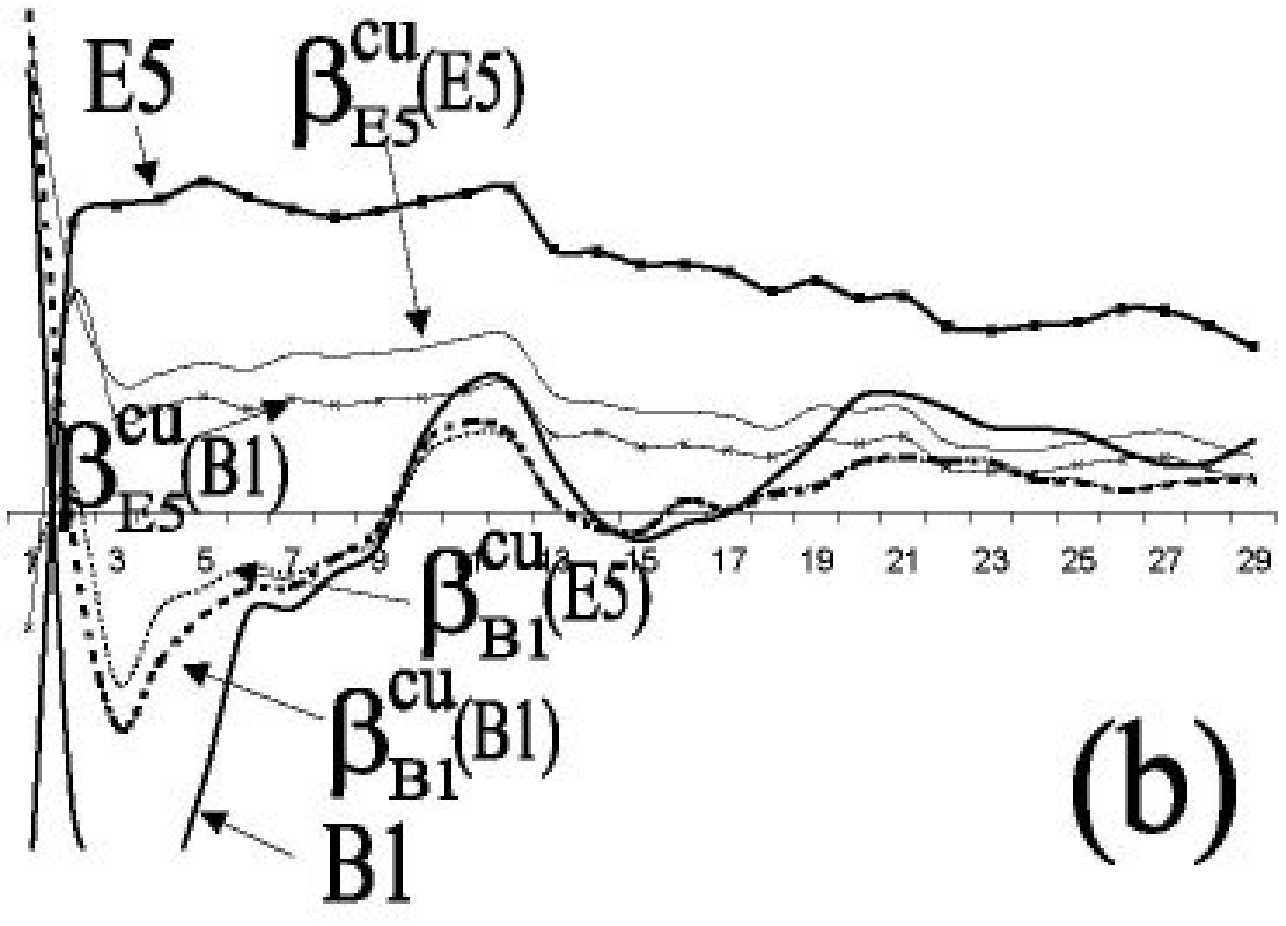}
  \end{center}
 \caption{
 (a) $\tilde{S}^5$ and (b) $\tilde{\Gamma}$ for the $\pi_{ws}$ projection of $B1$,
 $E5$, $\beta^{cu}_{B1}(B1)$, $\beta^{cu}_{B1}(E5)$, $\beta^{cu}_{E5}(B1)$ and
 $\beta^{cu}_{E5}(E5)$.}
 \label{fig:gr5}
 \end{figure}

 \subsection{Genetic Algorithms for Backtranslation}

  We want to obtain a backtranslation that imitates the genomic style of a target
  species as close as possible; thus, we will look for a backtranslation for which the
  coding statistics listed above are close to those of the target species, i.e., their
  distance is minimum. We choose, for $w\in B^{3*}$,
  \[
  \begin{array}{c}
   f_1(w) \;=\; |\rho_{g+c}(w)-\rho^*_{g+c}|
	\qquad 
	f_2(w) \;=\; \sum_{C\in B^3} |RSCU_C(w)-RSCU_C^*|   \\
	f_3(w) \;=\; |d_{ry}(w)-d_{ry}^*|+|d_{ws}(w)-d_{ws}^*|+|d_{mk}(w)-d_{mk}^*|  \\
	f_4(w) \;=\; \sum_{k=3}^{125} a_k |\tilde{S}^5_k(w)-\tilde{S}^{5,*}_k|
	\qquad 
	f_5(w) \;=\; \sum_{d=2}^{99} b_k |\tilde{\Gamma}_k(w)-\tilde{\Gamma}^*_k|
	\end{array} 
  \]
  where the values with ``*'' are obtained averaging over the known coding sequences
  of the target species, and $a_k$ and $b_k$ are weights, incorporated in order to
  give more importance to some parts of the curves, e.g. to encourage a uniform
  convergence. The indices in the sums of $\tilde{S}$ and $\tilde{\Gamma}$ follow
  our particular choices of window lengths 256 and 30, respectively. 

  With these definitions, what we want, for a given $u\in A^*$ and a given
  target species, is to minimize $\vec{f}(w)$, with $w\in \tau^{-1}(u)$.
  There are two main difficulties involved. First, we have a non-convex problem,
  in a vast search space, with terms depending on several scales of the
  sequences. Moreover, it is a problem of multiobjective optimization. For these
  reasons, we propose the use of genetic algorithms\cite{holland75} (GA), specially suited for 
  problems with these characteristics. Our particular implementation of a genetic
  algorithm for backtranslation follows here. 
  {\tt
   \begin{itemize}
	 \item
	  for $1\leq i\leq n$ initialize $w^i = \beta^{cu}(u)$
	 \item while not {\em stop condition}
	  \begin{itemize}
		 \item 
		  for $1\leq j\leq 5$, $\bar{f}_j = \max_{i} f_j(w^i)$
		 \item
		  for $1\leq i\leq n$, $1\leq j\leq 5$, $\displaystyle 
		  N_j^i = \frac{\bar{f}_j-f_j(w^i)}{n \bar{f}_j - \sum_k f_j(w^k)}$
		 \item
		  for $1\leq i\leq n$, $N^i = \sum_j \lambda_j N_j^i$
		 \item
		  Update $P$ using $\{N^i\}$ [stoch. univ. sampling]
		 \item
		  Apply genetic operators: crossover and mutation
		\end{itemize}
	\end{itemize}
  }
  For a given $u\in A^*$, we iterate on a population of $n$ guesses of
  $\tau^{-1}(u)$, denoted by $\{w^i\}$.
  As seen in the scheme, our initial condition is the usual backtranslation (imitation
  of RSCU); the GA is iterated then to optimize coding statistics.
  $N_j^i$ are the expected number of copies of a guess in the next generation;
  ponderating them with $\{\lambda_j\}$ we combine the different objective
  functions, without needing to make their numeric values comparable. The genetic
  operators used are crossover and mutation, both adapted to maintain the encoded
  amino acid sequence $u$. In addition, the probability of crossover between
  two guesses $w^i$ and $w^j$ depends on the Hamming distance between 
  them, making crossing between distant guesses less probable (this is introduced
  in order to encourage the exploration of a bigger region in search space).

  A special feature of this approximation is the use of the candidate solutions
  (guesses) as their own encodings for the GA. Of course, this is
  made possible by the sequential and digital nature of genetic sequences,
  which were the very inspiration of GA and other forms of evolutionary
  computation. Obvious as it may seem, this is the only application we know
  about in which genetic algorithms are applied to genetic sequences.

  \subsection{Results of GA application}
	\label{subsec:homol}
  
  The genetic algorithm was run several times for randomly selected sequences
  of $B1$ and $E5$ (with the other species as target, in each case), in order to find the
  best values for its parameters (mutation and crossover rates, population
  size, etc.), for the ponderations, etc.; this was done first for each $f_i$, and then for the
  combined optimization (detailed data can be found at \cite{web}). 
  Even when a single function was optimized, we
  computed all the statistics on the resulting guesses, in order to
  see the effect of each statistics on the rest. Optimization
  of spectra and autocorrelation functions, for instance, do not
  have the same effect on the sequence, in spite of working with the
  same information. Optimization of $\tilde{S}$ causes strong oscillations
  in $\tilde{\Gamma}$, whereas optimization of $\tilde{\Gamma}$ alone tends
  to cause a flattening of $\tilde{S}$. In general, imitation of $\tilde{\Gamma}$ 
  is the most difficult, followed by $\tilde{S}$, with $\rho_{g+c}$, $RSCU$ and
  specially IDH being the easier. The joint optimization of the $f_i$ arrived
  at values of each $f_i$ only slightly worse than those obtained in
  single function optimization, with the exception of $f_4$, which was
  actually better. Optimization of $\rho_{g+c}$ and $RSCU$ appeared to
  be almost unnecessary: when only $f_3$, $f_4$ and $f_5$ were considered
  (with $\beta^{cu}$ as initial condition), the final $\rho_{g+c}$ and
  $RSCU$ were still closer to the target species than the original sequence
  was to its own. In general, all $f_i$ {\em are} optimized by the genetic algorithm;
  it is even possible to make the periodicity of period 10.5 appear in
  sequences from which it was absent.

 \begin{figure}[hbt!]
  \begin{center}
   \includegraphics[width=15pc]{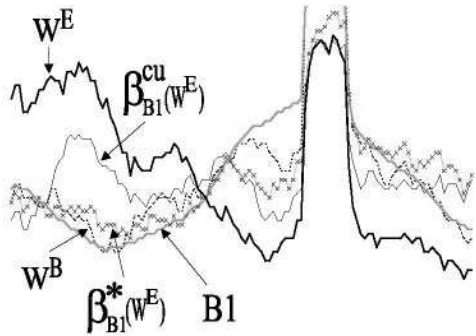}
  \end{center}
 \caption{
 $\tilde{S}^{11}$ for the $\pi_{ry}$ projection of $B1$, $W^B$, $\beta^{cu}_{B1}(W^E)$,
 $\beta^*_{B1}(W^E)$ and $W^E$.}
 \label{fig:gr6}
 \end{figure}

  To remove the differences due to the amino acid sequences (which can strongly
  influence any coding statistic in a sample with just a few sequences), we constructed
  a test set with sequences encoding homologue proteins in $B1$ and $E5$. To do this, we
  extracted from the {\em euGenes} database \cite{eugenes} the list of homologies
  between these species, chose the cases with a higher identity percentage, and cut
  the segment of each sequence corresponding to the alignment. Thus we obtained
  a set $W^B=\{w^B_1,\ldots,w^B_{20}\}$ of sequences from $B1$, and another set
  $W^E=\{w^E_1,\ldots,w^E_{20}\}$ from $E5$, with each pair $w^B_i, w^E_i$ encoding
  very similar amino acid sequences. We performed a canonical backtranslation
  on $\tau(W^E)$, obtaining $\beta^{cu}_{B1}(W^E)$; we perform also a backtranslation
  by means of our genetic algorithm, obtaining what we will call $\beta^*_{B1}(W^E)$.
  The computation of the diverse coding statistics allows us to see how this procedure
  gets the backtranslation closer to the average style of $B1$; moreover, since
  we do have $W^B$, we can compare with the values of that particular set of
  $B1$. For instance, for IDH, we can compute a distance between two
  sets of sequences $S_1$ and $S_2$ as $d_{idh}(S_1,S_2)=|d_{ry}(S_1)-d_{ry}(S_2)| 
  + |d_{ws}(S_1)-d_{ws}(S_2)| + |d_{mk}(S_1)-d_{mk}(S_2)|$. We obtain that
  $d_{idh}(W^E,W^B)=0.275$, while $d_{idh}(\beta^{cu}_{B1}(W^E),W^B)=0.104$, 
  and $d_{idh}(\beta^*_{B1}(W^E),W^B)=0.049$. Something similar happens with
  the other statistics. Graphic 6 shows the graphs of $\tilde{S}^{11}\circ \pi_{ry}$
  for the different sets; we can see again how $\beta^*$ builds a preimage for the
  image of $W^E$ (which is a typical $E5$ subset) which is far more similar to $B1$
  and $W^B$ than the usual backtranslation procedure, $\beta^{cu}$. 
  For $\tilde{\Gamma}$ the results are similar, but not so easy to observe in the
  graphics; instead of that, Table 3 displays the average 
  difference between the curve $\tilde{\Gamma}(W^B)$, and those for
  $W^E$, $\beta^{cu}_{B1}(W^E)$ and $\beta^*_{B1}(W^E)$. Again,
  $\beta^*$ improves with respect to $\beta^{cu}$.
  \small
\begin{center}
{\small 
 Table 3: Average distance of curves $\tilde{\Gamma}$} \\
 \begin{tabular}{|c|ccc|}\hline
 Projection & $d(W^E,W^B)$ & $d(\beta^{cu}_{B1}(W^E),W^B)$ & $d(\beta^*_{B1}(W^E),W^B)$ \\ \hline
 $\pi_{ws}$ & 0.0018 & 0.0013 & 0.0008 \\
 $\pi_{ry}$ & 0.0016 & 0.0019 & 0.0011 \\
    \hline
 \end{tabular}
 \end{center}
 \normalsize

 \section{Discussion}
\label{sec:lateral}

  The purpose of this article is to propose an improvement of the current 
  procedures of protein backtranslation, through the inclusion of 
  coding statistics other than RSCU which contribute to characterize the different genomes;
  this can be accomplished by the use of genetic algorithms. We first
  presented several known coding statistics, showing their idiosyncratic
  nature. Then we proposed a particular implementation of genetic algorithms,
  for a small set of coding statistics; this is only an example, since
  other choices of the statistics, or other implementations of evolutionary
  computation, may give better results. Our implementation, which is
  available at \cite{web}, {\em does} already produce backtranslations which mimic
  the coding statistics of the target species, in ways that are not automatically
  reproduced by RSCU imitation. 

  The definitive test for our approach would
  be the use of our procedure for the {\em in vitro} generation of actual 
  artificial genes: we expect it to have a higher success frequency than
  the canonical backtranslation. Meanwhile, the {\em in silico} experiment
  consisting in the backtranslation of a human protein into ``bacterial''
  style, and the comparison of the statistics of the resulting gene to those
  of an homologue bacterial gene (see section \ref{subsec:homol}), suggest that
  our approach is correct. In fact, the ``optimized'' preimages had
  more exact matches with the bacterial genes (at the aligned codon positions)
  than the simple RSCU-based backtranslation; this happened when human proteins
  were optimized for ``bacterial style'', and also when bacterial proteins
  were translated into ``human''. Though small, the systematic increase in exact matches
  is surprising: we did not expect the imitation of coding statistics to have this
  effect, since the number of preimages satisfying a given profile is still huge.

  This increase in exact matches suggests that the algorithm could be also
  applied to the problem of ``gene fishing'' through PCR reactions primed
  by degenerate primers, or ``guessmers''. This is a particular case of
  backtranslation, limited to short sequences selected for their minimal
  ambiguity. Thus, coding statistics are hard to evaluate (sequences
  are short) and hard to optimize (sequences are rigid). In spite of
  these difficulties, preliminary {\em in silico} experiments seem to support
  this application.

  Another field of application for the ideas presented here is the analysis
  of sequences: discussions on the relations and origins of coding statistics
  can be illuminated by massive backtranslation of sequences under some
  criteria, like we did in \ref{sec:relaciones} with RSCU to study its relation
  to spectra and autocorrelation functions. Of special interest are the
  comparisons between genes suspected, or known, to be related by horizontal
  transfer\cite{syvanen98}. Values of RSCU and/or $\rho_{g+c}$ divergent from
  the style of a genome have been used to detect horizontally transferred genes;
  the degree of their divergence has been used as a clock to determine when
  a gene was acquired\cite{sueoka93}. Some authors\cite{lawrence97} have done this through
  a ``reverse amelioration'' which is a kind of backtranslation, and could
  be enriched by the results and procedures given here.

%  The performance of the algorithm may be improved in several ways. If the
%  user knows something about the $\rho_{g+c}$ value of the particular region
%  were the gene is to be inserted, then it is possible to use only a subset
%  of the species' sequences as the ``target species'' (those with a similar
%  $\rho_{g+c}$ value). If the gene is expected to be highly expressed, the codon
%  usage information may be changed, to exagerate the codon bias. 
%  
%On the other hand, some improvements are still to be made to the code; the 
%preliminary successes with guessmer determination, and the local nature of most of the
%properties, suggest that a segmentation of the sequences, followed by 
% Otra posibles mejora: - incorporar efecto vecinos
%   
 \section{Acknowledgments}
   
	This research was supported by CONICYT through the FONDAP program in 
	Applied Mathematics, and was started during a visit to the 
	GREG ({\em Group de Recherche et d'Etude sur les Genomes}) at the Institut 
	of Mathematics at Luminy, University of Marseille, France. Special thanks
	to Alejandro Maass for his lasting support.

\end{document}